\documentclass[aip,apl,preprint]{revtex4-1}

\usepackage{subfigure}
\usepackage{graphicx}
\usepackage{eqnarray}
\usepackage{braket}
\usepackage{graphicx}
\usepackage{natbib}
\usepackage{url}
\usepackage{amsmath}
\usepackage{setspace}
\usepackage{siunitx}
\usepackage{longtable}
\usepackage{mciteplus}
\usepackage{epstopdf}
\usepackage{mhchem}
\usepackage{dsfont}

\begin{document}

\title{Nano-optical observation of cascade switching in a parallel superconducting nanowire single photon detector}

\author{Robert M. Heath} %
\email{r.heath.1@research.gla.ac.uk}
\author{Michael G. Tanner} %
\author{Alessandro Casaburi} %
\affiliation{School of Engineering, University of Glasgow, Glasgow, G12 8LT, Scotland}

\author{Mark G. Webster} %
\affiliation{Department of Statistics, University of Leeds, Leeds, LS2 9JT, England}

\author{Lara San Emeterio Alvarez} %
\author{Weitao Jiang} %
\author{Zoe H. Barber} %
\affiliation{Department of Materials Science and Metallurgy, University of Cambridge, Cambridge, CB3 0FS, England}

\author{Richard J. Warburton} %
\affiliation{Department of Physics, University of Basel, Klingelbergstrasse 82, CH-4056 Basel, Switzerland}

\author{Robert H. Hadfield} %
\affiliation{School of Engineering, University of Glasgow, Glasgow, G12 8LT, Scotland}

\begin{abstract}
The device physics of parallel-wire superconducting nanowire single photon detectors is based on a cascade process. Using nano-optical techniques and a parallel wire device with spatially-separate pixels we explicitly demonstrate the single- and multi-photon triggering regimes. We develop a model for describing efficiency of a detector operating in the arm-trigger regime. We investigate the timing response of the detector when illuminating a single pixel and two pixels. We see a change in the active area of the detector between the two regimes and find the two-pixel trigger regime to have a faster timing response than the one-pixel regime.
\end{abstract}

\maketitle

Superconducting nanowire single photon detectors\cite{gol2001picosecond} (SNSPDs) are considered exceptionally fast single photon detectors with low jitter\cite{natarajan2012superconducting} and operate over a large spectral range, from visible to mid-infrared\cite{marsili2012efficient}. This makes them desirable for applications in satellite communications\cite{robinson2006781}, medical physics\cite{gemmell2013singlet}, fiber-optic distributed Raman temperature measurement\cite{tanner2011high}, and quantum communications and computing\cite{obrien2007optical,bonneau2012fast}. For observing photon anti-bunching in quantum optics and in many areas of quantum information\cite{gisin2002quantum} photon number resolution is desirable\cite{sahin2013waveguide}: a single-wire SNSPD does not have this property.

Parallel-wire variants of SNSPDs\cite{ejrnaes2007cascade} offer the possibility of higher count rates\cite{ejrnaes2011characterization}, thinner wires with higher signal-to-noise\cite{najafi2012timing}, and multi-photon detection. These devices, also referred to as `cascade-switching' detectors or `superconducting nanowire avalanche photodetectors'\cite{najafi2012timing}, are able to operate in a different regime to single-wire SNSPDs, triggering only on the absorption of at least two photons\cite{marsili2010single}. Typical geometries employed to date when patterning these devices have 2--24 spatially-indistinguishable, often interleaved nanowires connected in parallel\cite{ejrnaes2011characterization,marsili2010single,najafi2012timing,ejrnaes2007cascade}. As a result of this layout, it is impossible to explicitly observe what happens when only one wire of the parallel structure is illuminated, and in turn distinguish one- and two-photon detection. In this work we have fabricated a device with spatially-separated multiple wires connected in parallel (referred to as pixels) to capture these distinct absorption regimes.

In single-wire SNSPD operation, a superconducting nanowire is biased below its critical current $I_{c}$ at bias current $I_{b}$. Photons striking the superconductor create a hotspot, which decreases the superconducting cross-section, pushes the current density above the superconducting threshold, and so in turn the full width of the nanowire is driven into the resistive state. This abrupt interruption in the current flow results in a fast voltage pulse as the detector absorbs a photon. In a parallel-wire device, $n$ wires are biased such that, if the bias current is sufficiently low\cite{marsili2011electrothermal} ($I_{b}<I_{av}$ where $I_{av}$ is the avalanche or cascade threshold current), one may absorb a photon without immediately triggering a second. If $I_{b}$ is sufficiently near $I_{c}$ ($I_{b}>I_{av}$), a single photon pushing a single wire resistive will trigger the cascade process\cite{ejrnaes2007cascade}. It is by operating in the arm-trigger regime that multi-photon discrimination is possible.

To observe the spatially-separate one- and two-photon absorption regimes on a multi-pixel device, one must have a sufficiently small spot of light, and some method of accurately translating this across the device. We have used a miniature confocal microscope with low temperature piezoelectric nanopositioners in previous work\cite{oconnor2011spatial,tanner2012superconducting} which we exploit again for this purpose. Work by others exploring the trigger mechanism of parallel-wire SNSPDs\cite{marsili2010single} under broad illumination shows that, in a multi-wire device biased at a low $I_{b}$, an arm-trigger mechanism occurs: the first photon will `arm' the device and the second will `trigger' it: unless the device has a trigger pulse as well, or is triggered by a dark-count, the detector will behave as though nothing was detected, and will ultimately reset as the hotspot energy dissipates into the substrate.

The device under test, a spatially-separate multi-pixel SNSPD fabricated in niobium nitride on a sapphire substrate with \SI{55}{\nano\metre}-wide wires at \SI{130}{\nano\metre} pitch, was cooled to \SI{3.5}{\kelvin} in a vibration-damped closed-cycle cryostat based on a pulse-tube refrigerator. The miniature confocal microscope\cite{oconnor2011spatial} was employed to realise photoresponse maps of the device, which had a film thickness of \SI{8}{\nano\metre} and a superconducting transition temperature of \SI{9}{\kelvin}. A critical current of \SI{21}{\micro\ampere} was measured when operated with a \SI{50}{\ohm} shunt resistor in parallel at \SI{3.5}{\kelvin}. Large meanders were patterned at either end of the active area to increase the kinetic inductance of the device which, along with the shunt resistor, allowed it to reset without latching.

\begin{figure}[]
	\setlength\fboxsep{0pt}
	\setlength\fboxrule{0.25pt}
	\begin{center}
		\includegraphics[width=0.48\textwidth]{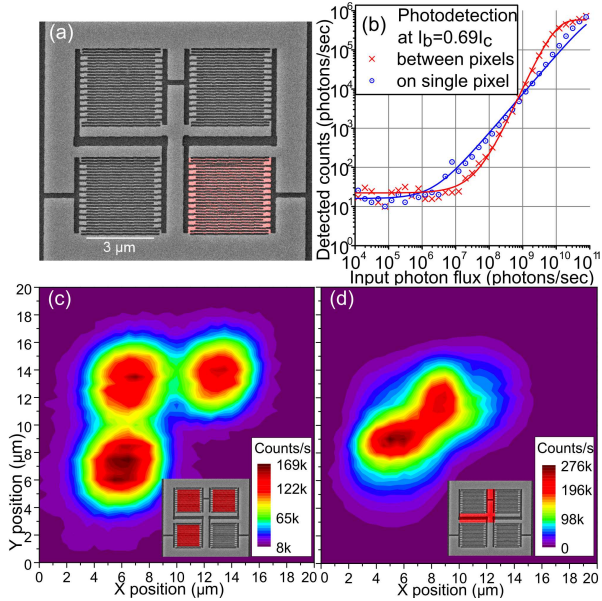}
	\end{center}
	\caption{Figure~\ref{fig:explanation-figure}(a) shows a scanning electron microscope (SEM) image of the detector, with NbN bright and substrate darker. Each pixel is spatially separate but electrically connected in parallel. One pixel (shown in red) failed---the broken wire can be seen clearly, therefore this is in effect a 3-pixel parallel nanowire device. Figure~\ref{fig:explanation-figure}(c) shows the high-bias ($0.83I_{c}$, $25$ photons per pulse) photoresponse with position of the three pixels that were successfully fabricated: it is clear that the lower pixel counts more than the other two, which we attribute to a small non-uniformity in film thickness during fabrication. In figure~\ref{fig:explanation-figure}(d) we see two-pixel discrimination highlighting the areas where there are two active pixels, which occurs at lower bias ($0.69I_{c}$, $8000$ photons per pulse). A small blue `halo' around the between-areas' response shows that the single pixels still respond in this regime, though at a much reduced rate.
	Figure~\ref{fig:explanation-figure}(b) shows count rate as a function of photon flux entering the system when illuminating one- and two-pixels at fixed bias current. Maximized-$R^{2}$ models of photon absorption in these regimes (equations~\ref{eqn:Rdet} and~\ref{eqn:Rdet2} respectively) were fitted. In the single-pixel illumination case, the fit has efficiency $\eta=7.5 \times 10^{-4}$, while the two-pixel illumination has efficiency $\eta=1.59\times 10^{-4}$. The efficiency is expected to fall in the two-pixel case, as much of the optical spot is illuminating the inactive area between pixels. The low bias point ($0.69 I_{c}$) was used to keep the dark-count rate constant, and shows that even at low bias, both one- and two-pixel triggering regimes are possible, albeit with low efficiency.} 
	\label{fig:explanation-figure}
\end{figure}

Light of wavelength \SI{1550}{\nano\metre} from a \SI{1}{\mega\hertz} pulsed diode laser was delivered by optical fiber into the closed-cycle cryocooler. The light was then collimated in a miniature confocal microscope and focused to a diffraction-limited spot with FWHM \SI{1.3}{\micro\metre} which is able to move relative to the detector. Large-range Attocube piezoelectric motors (\SI{5}{\milli\metre} range) allow rough placement of the spot, while a piezoelectric X-Y scanner (\SI{40}{\micro\metre}$\times$\SI{40}{\micro\metre} scan area at \SI{3.5}{\kelvin}, sub-micrometer precision) allows accurate characterisation of the detector response.

To allow readout from the detector for the counts maps in figure~\ref{fig:explanation-figure}, the device was biased by a voltage source in series with a \SI{100}{\kilo\ohm} resistor and low-pass filter through a bias tee's DC `input'. The RF + DC side was shunted with a \SI{50}{\ohm} resistor to allow the detector to reset and connected to the SNSPD. The RF output of the circuit was amplified (gain \SI{56}{\deci\bel} typical, amplifier bandwidth \SI{10}{\mega\hertz}--\SI{580}{\mega\hertz}) and connected to a counter.

The scanning electron microscope (SEM) image in figure~\ref{fig:explanation-figure}(a) shows the layout of the device. Each of the four parallel pixels is $\SI{3}{\micro\metre}\times\SI{3}{\micro\metre}$, though the bottom-right pixel has a broken wire, meaning the device behaves as a 3-pixel device. The photoresponse of the active pixels is shown for high bias ($0.83 I_{c}$) in figure~\ref{fig:explanation-figure}(c) and for low bias ($0.69 I_{c}$) in figure~\ref{fig:explanation-figure}(d). These plots suggest that the pixels are not perfectly uniform, with the lower pixel approximately \SI{10}{\percent} more responsive than the other two. The individual pixels respond at high bias, while at low bias and high flux the areas between multiple pixels respond most strongly: this agrees with the concept of an arm-trigger regime at lower bias\cite{marsili2010single}.

A detector with a parallel configuration requires a model of how it detects photons in its various operating regimes. Equation~\ref{eqn:Rdet} is the description of the detection behavior in the single-pixel case for the count rate $R_{detected}$ of a detector illuminated by a pulsed laser\cite{hadfield2009single}
\begin{equation}
\label{eqn:Rdet}
R_{detected}= f \left( 1 - (1-\eta)^\mu \right) \approx f \left(1-e^{-\mu\eta} \right),
\end{equation}
where $f$ is laser pulse frequency, $\mu$ is mean photon number per pulse, and $\eta$ is the detector efficiency. 

To describe multi-pixel triggering we assume the detector only outputs a pulse if $b$ of the $n$ pixels are triggered ($b$ depends on the bias current and $n$ is defined by the detector geometry), and the active area of the detector changes as each pixel arms. We assume even illumination of the pixels, which is valid if the optical spot is illuminating between pixels. We must first consider dark counts: one may arm the device, meaning a single photon can trigger the detector even in the two-pixel regime, or two must coincide on separate pixels for the counter to register them. For this reason we consider the switching of individual pixels without a photon being absorbed as a `dark event' - not all dark events result in a registered count. Modelling the dark event rate as a Poisson process with rate $\lambda$ gives probability $\gamma=1-e^{-\lambda}$. In our model, the total number of dark events $g$ affect the trigger probability, and have a binomial distribution described as \begin{equation}
\label{eqn:DC}
p(g|\gamma)={n \choose g} \gamma^g \left(1-\gamma\right) ^{n-g}.
\end{equation}
Considering then the $\mu$ incident photons, we have $\eta$ probability of each to form a hotspot, which gives $d$ detectable (though doesn't tell us in which pixels the arming events happen; they could all occur in the same pixel, in which case there would be no trigger). Taking the dark events into account, the chance of having $d$ arming events is binomial, equal to
\begin{equation}
\label{eqn:binom}
p(d|\mu,\eta,g)={\mu \choose d} \eta^d \left(1-\frac{n-g}{n}\eta\right) ^{\mu-d}.
\end{equation}

The probability of whether a triggering event $T$ occurs  $\mathds{P}(T|\mu,\eta,\gamma)$ is formed by multiplying the probability of having a certain number of arming events by the probability of those events being distributed in such a way that at least $b$ pixels are armed, such that  $\mathds{P}(T|\mu,\eta,\gamma) = \sum\limits_{d=0}^{\mu}p(d|\mu,\eta,\gamma)\mathds{P}(T|d,\gamma)$. Combining this with the pixels armed by dark counts gives
\begin{equation}
\label{eqn:general}
\mathds{P}(T|\mu,\eta,\gamma) = \sum\limits_{g=0}^{b}p(g|\gamma)\sum\limits_{d=0}^{\mu}p(d|\mu,\eta,g)\mathds{P}(T|d,g).
\end{equation}
Solving equation~\ref{eqn:general} is non-trivial for general $b$. However, when two pixels are required to cascade ($b=2$) as in this case, and dark events $g=0$, $\mathds{P}(T|d,g)=(1-n^{1-d})(1-\delta_{d,0})$ where $\delta_{a,b}$ is the Kronecker delta, equal to one if $a=b$, else zero. When $g=1$, $\mathds{P}(T|d,g)=(1-\delta_{d,0})$, and when $g\geq 2$, $\mathds{P}(T|d,g)=1$. In this case, by summing from $d=1$, the $g=0$ case becomes $\mathds{P}(T|d,g)=(1-n^{1-d})$ and the $g=1$ case becomes $1$, and so with two pixels illuminated equally we have
\begin{align}
\mathds{P}(T|\mu,\eta,\gamma)
=&\sum_{g=0}^2 {2 \choose g} \gamma^g \left(1-\gamma\right) ^{2-g} \sum_{d=2-g}^{\mu}{\mu \choose d} \left(\frac{2-g} {2} \eta\right)^d \left(1-\frac{2-g} {2} \eta\right) ^{\mu-d} \mathds{P}(T|d,g) \nonumber \\
=&(1-\gamma)^2 \sum_{d=2}^{\mu}{\mu \choose d} \eta^d \left(1-\eta\right) ^{\mu-d} (1-2^{1-d} ) \nonumber \\
&+2\gamma(1-\gamma) \sum_{d=1}^{\mu}{\mu \choose d} \left(\frac{\eta} {2} \right)^d \left(1-\eta/2\right) ^{\mu-d} +\gamma^2 \nonumber \\
=&1+(1-\gamma)^2 (1-\eta)^{\mu} -2(1-\gamma) (1-\eta/2)^{\mu}, \nonumber
\end{align}
which can be restated as equation~\ref{eqn:Rdet2}. This is used to fit the two-pixel data in figure~\ref{fig:explanation-figure}(b).
\begin{equation}
\label{eqn:Rdet2}
R_{detected}=f \left(1+ \left(1-\gamma\right)^{2}\left(1-\eta \right)^\mu - 2 \left(1-\gamma\right)\left(1-\eta/2\right)^\mu \right)
\end{equation}
This model does not allow for pixels armed in a previous pulse window (`pre-armed' pixels), which is a possibility when operating in the two-pixel trigger regime: the current redistribution\cite{casaburi2013current} suggests that the unarmed pixels would experience a much higher relative bias while the pre-armed pixel supports little. However, dark events increase with bias, which would thus increase the probability of another pixel triggering before the next pulse and resetting the detector. Pre-arming the detector would give a linear dependence on photon flux where we see a quadratic, suggesting that this effect does not contribute significantly. Current redistribution may reset the detector between pulse windows, resulting in an increased dark count rate. 

With two pixels but when only one is required to trigger ($n=2$, $b=1$, $g=0$), $\mathds{P}(T|d)=1-\delta_{d,0}$ which by summing from $d=1$ simplifies to $\mathds{P}(T|d)=1$. This leads to equation~\ref{eqn:Rdet3}
\begin{align}
\label{eqn:Rdet3}
\mathds{P} (T|\mu,\eta,\gamma)
&=\sum_{g=0}^2 p(g|\gamma) \sum_{d=0}^{\mu} \binom{\mu} {d} \left(\frac{2-g} {2} \eta\right)^d \left(1-\frac{2-g} {2} \eta\right)^{\mu-d} \mathds{P} (T|d,g) \nonumber \\
&=(1-\gamma)^2 \sum_{d=1}^{\mu} \binom{\mu} {d} \eta^d (1-\eta)^{\mu-d} +2\gamma(1-\gamma) +\gamma^2 \nonumber \\
&=1-(1-\gamma)^2 (1-\eta)^{\mu},
\end{align}
and so, if we do not take into accounts the dark events ($\gamma=0$) we see that the existing description of single-pixel triggering in equation~\ref{eqn:Rdet} can be derived. This may be used to fit single-pixel data. As is shown in figure~\ref{fig:explanation-figure}(b), these equations fit well: the two-pixel case is much steeper, which is to be expected as it has a greater dependence on photon flux. The SNSPD transitions seamlessly between the two regimes as the bias current $I_{b}$ is varied. At $I_b=0.69I_c$ there is clearly both one- and two-pixel triggering occurring. When operating in the between-pixel illumination regime the device will occasionally trigger on a single pixel, especially at low flux when two-pixel triggering is unlikely. When illuminating on-pixel, some of the light some of the time will form a hotspot on a second pixel and a cascade will happen via the two-pixel process. This effect will cause the data to deviate from the theory, which only describes a single regime at a time.

\begin{figure}[]
	\begin{center}
	    \includegraphics[width=0.48\textwidth]{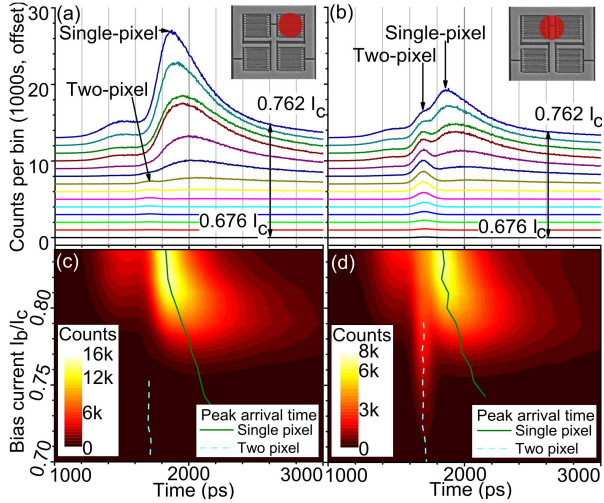}
	\end{center}
	\caption{The photon arrival time profiles of two cases: figures~\ref{fig:illumination-profile}(a) and~\ref{fig:illumination-profile}(c) illuminating a single element, and figures~\ref{fig:illumination-profile}(b) and~\ref{fig:illumination-profile}(d) illuminating two elements. The stacked line section plots are offset in Y to make them more visually distinguishable. As the bias current is reduced in the one-pixel case, the device triggering delay increases, until it triggers no further and the response drops away. In the two-pixel illumination case, the device behaves similar to the one pixel case at high bias, but as the bias is reduced, a second signal peak emerges, ahead of the single photon absorption peak and with comparable FWHM jitter. In this regime we are discriminating two-photon events. The peak arrival times for single- and two-pixel triggering are shown on figures~\ref{fig:illumination-profile}(c) and~\ref{fig:illumination-profile}(d) for one-pixel trigger (green line) and two-pixel trigger (cyan dash). As the bias decreases and the jitter increases and arrival times are delayed, the count rate of the detectors decreases, which can be seen in figure~\ref{fig:illumination-profile}(c) and figure~\ref{fig:illumination-profile}(d).}
	\label{fig:illumination-profile}
\end{figure}

In figure~\ref{fig:illumination-profile}, we present the timing properties of the device over different bias and illumination conditions. A time-correlated single photon counting (TCSPC) technique was employed and the device was current-biased through a bias tee, with a \SI{50}{\ohm} shunt resistor. The amplified output was run to the input of a PicoQuant PicoHarp 300. The input light at wavelength \SI{1550}{\nano\metre} from a Kphotonics CNT-1550 mode-locked erbium-doped fiber laser was pulsed at \SI{50}{\mega\hertz} with picosecond pulse width, and was split between an \ce{InGaAs} fast photodiode (intrinsic jitter \SI{15}{\pico\second}) and the parallel-wire SNSPD. The fast photodiode was connected to the reference input of the PicoHarp. This allows the PicoHarp to measure, in \SI{4}{\pico\second} bins, the temporal response of the detector with respect to the reference pulse, from which we can obtain the jitter and peak arrival times.

Figure~\ref{fig:illumination-profile}(a), taken illuminating one pixel, shows line plots of the response of the parallel-wire device when one pixel is illuminated as its bias is reduced from its single-photon-sensitive regime. At the peak, the number of counts decreases, and the position is delayed, with the characteristic `tail' explained by Ejrnaes\cite{ejrnaes2009timing} that others also observe\cite{najafi2012timing,marsili2010single} in the instrumental response of parallel wire devices becoming more pronounced. This is also shown in figure~\ref{fig:illumination-profile}(c), which more clearly shows the tail.

Comparing figure~\ref{fig:illumination-profile}(a) with figure~\ref{fig:illumination-profile}(b) which shows the response when two pixels are illuminated, an additional feature is seen---a small bump on the rising edge of the response time peak. As the bias decreases in the two-pixel regime, the dominant peak (single photon response) reduces significantly, while the bump, the two-photon response, becomes stronger until it is more prominent than the single photon response. This is seen in figure~\ref{fig:illumination-profile}(d) as protrusion from the single-photon response area, maintaining high count rates per bin to much lower bias than in figure~\ref{fig:illumination-profile}(a). The reduction of the count rate from the on-pixel illumination to the two-pixel is due to the change in active area with respect to the center of the optical spot, while the increased count rates at low bias in the between-pixel illumination regime is due to the absorption of two or more photons on at least two pixels, enabling the cascade process.

Examining the timing properties of the device using TCSPC, on figures~\ref{fig:illumination-profile}(c) and~d the peak arrival times are displayed. For both one- and two-pixel illumination the `single photon' relative peak arrival time is lowest at high bias, and the full-width half-maximum (FWHM) jitter of this peak is narrow, suggesting the device consistently responds quickly. As the bias is decreased, the time to the peak is delayed, and the jitter becomes larger. The change in relative peak arrival time is greater than other results in the field\cite{najafi2012timing}, though none of these feature long meanders and our TCSPC equipment, contrary to some, measures to the peak of the pulse. These factors both independently increase this value. Between $0.78I_{c}$ and $0.74I_{c}$ for between-pixel illumination there is an overlap in one- and two-pixel triggering regimes, seen in the growing left peak in figure~\ref{fig:illumination-profile}(b) and that there are both one- and two-pixel peak arrival times in figure~\ref{fig:illumination-profile}(d). The lack of a distinct switch from one regime to another may be due to the non-uniformity of the device. Even so, these values tie in with electrothermal simulation by Marsili\cite{marsili2011electrothermal}, which shows that a uniform device with three wires in similar layout should move to the two-pixel trigger regime at $0.78I_{c}$. Two wires simultaneously arming will divert current to the remaining wire at a faster rate, resulting in the faster relative peak arrival time seen in the `two photon' case. 

When illuminating one pixel, some small number of photons are able to strike a second pixel which results in there being a small two-photon response which emerges at $~0.76I_c$. This is significant enough to show in figure~\ref{fig:illumination-profile}(a), though as is clear from its profile it occurs with very low count rates, and is included for completeness. As the bias is reduced, both the single- and two-photon responses are delayed, as shown in figures~\ref{fig:illumination-profile}(c) and~d. We also see from the width of the profiles in figures~\ref{fig:illumination-profile}(a) and~\ref{fig:illumination-profile}(b) there is an increase in the FWHM jitter. A simple electrical model shows that at lower bias, it takes longer for sufficient current to be diverted to the remaining wire for a cascade to take place.

Understanding the device physics of parallel SNSPDs opens the pathway to large-area parallel SNSPDs with extended spectral range, low jitter, short reset times and multi-photon discrimination. Using a low-temperature nano-optical measurement system we present a spatially-distinguishable demonstration of the one- and two-pixel trigger regimes of parallel-wire SNSPDs. We have formulated a theoretical description of the efficiency of these devices, and shown that while it is less efficient in the multi-pixel trigger regime owing to the reduced active area caused by `arming' a single pixel, the timing properties show the device to be faster than when it is operating in single-photon detection mode. This work explicitly demonstrates and describes the temporal features of a spatially-separate parallel-wire SNSPD applicable to short-length and interleaved designs. 

The authors thank the UK Engineering and Physical Sciences Research Council for support. RHH acknowledges a Royal Society of London University Research Fellowship and EPSRC grants EP/I036273/1 and EP/G022151/1. ZHB acknowledges EPSRC grant EP/G022208/1. AC acknowledges a Marie Curie Fellowship. RJW acknowledges support from NCCR QSIT.

\end{document}